PAPER • OPEN ACCESS

# Higher first Chern numbers in one-dimensional Bose–Fermi mixtures

To cite this article: Kristian Knakkergaard Nielsen et al 2018 New J. Phys. 20 025005

View the article online for updates and enhancements.

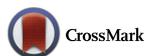

PAPER

OPEN ACCESS

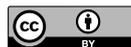

# Higher first Chern numbers in one-dimensional Bose–Fermi mixtures

Kristian Knakkergaard Nielsen[1], Zhigang Wu[2] and G M Bruun[1]

[1] Department of Physics and Astronomy, Aarhus University, Ny Munkegade, DK-8000 Aarhus C, Denmark
[2] Institute for Advanced Study, Tsinghua University, Beijing, 100084, People's Republic of China

E-mail: bruungmb@phys.au.dk

## Abstract

We propose to use a one-dimensional system consisting of identical fermions in a periodically driven lattice immersed in a Bose gas, to realise topological superfluid phases with Chern numbers larger than 1. The bosons mediate an attractive induced interaction between the fermions, and we derive a simple formula to analyse the topological properties of the resulting pairing. When the coherence length of the bosons is large compared to the lattice spacing and there is a significant next-nearest neighbour hopping for the fermions, the system can realise a superfluid with Chern number $\pm 2$. We show that this phase is stable in a large region of the phase diagram as a function of the filling fraction of the fermions and the coherence length of the bosons. Cold atomic gases offer the possibility to realise the proposed system using well-known experimental techniques.

## 1. Introduction

Until fairly recently, only a few topological phases were known to exist in nature [1–4]. We now know that these phases are surprisingly ubiquitous in nature, and their study has consequently witnessed an explosion of activities [5–7]. Topological superfluids[3] are particularly interesting, since they support gapless edge states, the so-called Majorana modes, which have possible applications for quantum computation [8, 9]. In a seminal paper, Kitaev introduced a one-dimensional (1D) model of spin-less fermions forming $p$-wave pairs [10] which has attracted enormous attention since it is one of the simplest playgrounds to study superfluids with non-trivial topological properties. There is an intense effort to observe these modes experimentally, and promising evidences for their existence have been reported in systems where a 1D wire is placed in proximity to a conventional superconductor [11–17]. The Kitaev wire has, in addition to a particle–hole symmetry, also a time-reversal symmetry which squares to be one. It follows that it belongs to class BDI in the classification scheme of topological insulators and superfluids. This class has a $\mathbb{Z}$ topological invariant in 1D [18–20], and it therefore allows the existence of multiple orthogonal Majorana edge states. In order to bring out this $\mathbb{Z}$ character of the topological invariant and the associated multiplicity of Majorana states, one however needs to generalise the Kitaev model to include long range pairing and hopping. This has been demonstrated in several models where the long range nature has been introduced by hand [21–23].

The aim of this paper is to show how the $\mathbb{Z}$ character of the topological invariant of a generalised Kitaev model can be realised in an experimentally realistic system consisting of identical fermions in a 1D lattice immersed in a 1D Bose gas. We calculate the induced interaction between the fermions caused by the exchange of density oscillations in the Bose gas. The topological properties of the resulting pairing between the fermions is then analysed by deriving a simple formula for the Chern number. We demonstrate that the system can realise a phase with Chern number $\pm 2$, when the coherence length of the bosons is large and the fermions has a significant next-nearest neighbour hopping. Furthermore, we show that this phase is stable in large regions of the phase diagram as a function of the filling fraction of the lattice and the coherence length of the bosons. Finally, we briefly discuss the experimental realisation of the proposed system using cold atomic mixtures in a periodically driven lattice.

---

[3] We refer to superfluids only from now on. It is implicitly understood that this also refers to superconductors.

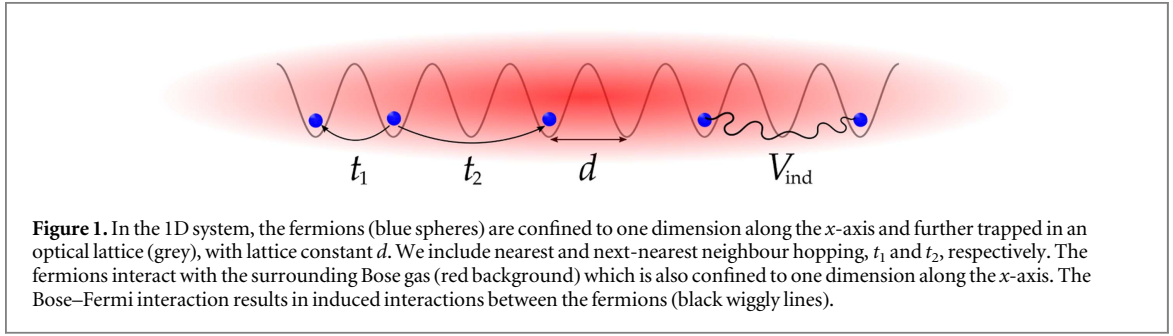

**Figure 1.** In the 1D system, the fermions (blue spheres) are confined to one dimension along the *x*-axis and further trapped in an optical lattice (grey), with lattice constant *d*. We include nearest and next-nearest neighbour hopping, $t_1$ and $t_2$, respectively. The fermions interact with the surrounding Bose gas (red background) which is also confined to one dimension along the *x*-axis. The Bose–Fermi interaction results in induced interactions between the fermions (black wiggly lines).

## 2. System

We consider identical (spin-polarised) non-interacting fermions of mass $m_F$ mixed with bosons of mass $m_B$ in a cigar-shaped trap. The transverse trapping frequency is much larger than any relevant energy scale, so that the system can be treated as 1D. The fermions are subjected to an additional periodic potential forming a 1D lattice whereas the bosons move freely in 1D as illustrated in figure 1. We further assume that the 1D lattice is sufficiently deep so that the fermions stay in the lowest energy band.

We take the *x*-axis to be along the 1D direction and densities of the fermions and bosons are $n_F$ and $n_B$, respectively. The Bose–Fermi interaction is modelled by a contact interaction $g_{BF}\delta(\mathbf{r})$ with a coupling strength $g_{BF}$. The Bose–Bose interaction is also of the form $g_B\delta(\mathbf{r})$, where $g_B$ is the Bose–Bose interaction strength, whereas the fermions do not interact with themselves. We describe the bosons using Bogoliubov theory, which has been shown to yield the correct excitation spectrum despite the fact that there is no true Bose–Einstein condensation in a 1D homogeneous system [24, 25]. The Hamiltonian is

$$H = -\sum_j [t_1(c^\dagger_{j+1}c_j + c^\dagger_j c_{j+1}) + t_2(c^\dagger_{j+2}c_j + c^\dagger_j c_{j+2})]$$
$$+ \sum_k E_k^B \gamma_k^\dagger \gamma_k + g_{BF}\int d^3r \psi_F^\dagger(\mathbf{r})\psi_B^\dagger(\mathbf{r})\psi_B(\mathbf{r})\psi_F(\mathbf{r}), \qquad (1)$$

where $c_j^\dagger$ creates a fermion in lattice site *j* with the nearest and next-nearest neighbour hopping $t_1$ and $t_2$, respectively, and $\gamma_k^\dagger$ creates a Bogoliubov mode with momentum *k* and energy $E_k^B = [\varepsilon_k^B(\varepsilon_k^B + 2\tilde{g}_B n_B)]^{1/2}$. Here $\varepsilon_k^B = k^2/2m_B$. We have defined $\tilde{g}_B = g_B/2\pi l_\perp^2$ as the effective boson–boson coupling in 1D, where $l_\perp$ is the harmonic oscillator length for the transverse trap. The Bose–Fermi interaction is expressed in terms of Bose and Fermi field operators $\psi_B(\mathbf{r})$ and $\psi_F(\mathbf{r})$ and will be simplified later. In a typical static optical lattice employed in current experiments, the next-nearest neighbour hopping $t_2$ is at least one order of magnitude smaller than $t_1$. However, by shaking the optical lattice one can increase the effective next-nearest neighbour hopping such that the ratio $t_2/t_1$ can be tuned. This technique has in fact been implemented by several experimental groups [26–31]. In momentum space, we can write the tight-binding part of the Hamiltonian for the fermions, i.e. the first line in (1), as $\sum_{k=-\pi/2d}^{\pi/d} \varepsilon_k c_k^\dagger c_k$ with $\varepsilon_k = -2t_1\cos kd - 2t_2\cos 2kd$. The appearance of the $\cos(2kd)$ term in the fermion dispersion will turn out to be one of the crucial ingredients to realise a phase with a Chern number larger than one.

To treat the Bose–Fermi interaction, we expand the fermion field operator as $\psi_F(\mathbf{r}) = \sum_j \phi_\perp(y,z) w(x-x_j) c_j$ where $\phi_\perp(y,z)$ is the ground state of the transverse harmonic trap and $w(x-x_j)$ is the lowest band Wannier function located at $x_j$ in the 1D optical lattice. Likewise, the boson field operator is expanded as $\psi_B(\mathbf{r}) = L^{-1/2}\sum_k \phi_\perp(y,z)\exp(ikx) b_k$, where *L* is the 1D length of the system and $b_k$ creates a boson with momentum *k*. Using these expansions, we can write the boson–fermion interaction, i.e. the third term in (1), as

$$H_{int} = \frac{g_{BF}}{2\pi l_\perp^2 L}\sum_j \sum_{k,q} e^{iqx_j} c_j^\dagger c_j b_{k+q}^\dagger b_k, \qquad (2)$$

where we have approximated the Wannier function by the ground state of the local harmonic potential and used $\int dx\, w^*(x-x_i)w(x-x_j)e^{iqx} \simeq 0$ with $i \neq j$ for well localised Wannier functions. To relate the $b_k$ operator to the $\gamma_k$ operator, we note that the Bogoliubov theory gives the usual relation $b_k = u_k\gamma_k - v_k\gamma_{-k}^\dagger$ with $u_k^2 = [1 + (\varepsilon_k^B + \tilde{g}_B n_B)/E_k^B]/2$ and $v_k^2 = u_k^2 - 1$.

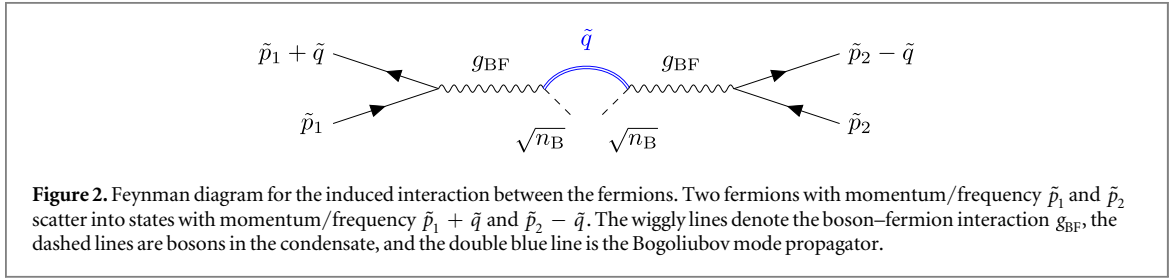

**Figure 2.** Feynman diagram for the induced interaction between the fermions. Two fermions with momentum/frequency $\tilde{p}_1$ and $\tilde{p}_2$ scatter into states with momentum/frequency $\tilde{p}_1 + \tilde{q}$ and $\tilde{p}_2 - \tilde{q}$. The wiggly lines denote the boson–fermion interaction $g_{BF}$, the dashed lines are bosons in the condensate, and the double blue line is the Bogoliubov mode propagator.

## 3. Induced interaction

Recently, two of us demonstrated that a 3D BEC is an effective medium to induce an attractive interaction between identical fermions, giving rise to a *p*-wave superfluid with a high critical temperature [32]. Here, we extend this idea to the 1D case. The fermions will interact with each other via the exchange of Bogoliubov modes in the BEC. For a weak Bose–Fermi interaction $n_F g_{BF}/2\pi l_\perp^2 \ll 1$, the leading order diagram is the one shown in figure 2.

The resulting induced interaction in real space is

$$V_{\text{ind}}(x, \omega) = \left(\frac{g_{BF}}{2\pi l_\perp^2}\right)^2 \int \frac{dq}{2\pi} e^{iqx} \chi_B(q, \omega), \quad (3)$$

where

$$\chi_B(q, \omega) = \frac{q^2}{m_B} \frac{n_B}{\omega^2 - (E_q^B)^2} \quad (4)$$

is the density–density correlation function for the bosons within Bogoliubov theory, and $\omega$ is the transferred energy between the two interacting fermions. It has been shown that this approximation for the density–density correlation function is accurate for a weakly interacting 1D Bose gas [33]. This justifies our use of Bogoliubov theory, since the bosons enter the theory only through their density–density correlation function. The frequency dependence of the induced interaction (3) reflects that the interaction is not instantaneous due to the finite speed of the Bogoliubov modes, which gives rise to retardation effects. It can be shown that these effects are small when the Fermi velocity $v_F$ of the fermions is much less than the speed of sound in the Bose gas, while for larger $v_F$ they suppress the magnitude of the pairing without changing the qualitative behaviour [32]. We therefore neglect retardation effects in the following and restrict ourselves to the $\omega = 0$ component of (3). This static induced interaction is evaluated in the following manner. First, we write the integral on unitless form as: $V_{\text{ind}}(x, 0) = G/\pi \int d\tilde{q} \ e^{i\tilde{q}\tilde{x}}/(\tilde{q}^2 + 1)$, with $\tilde{q} = q\xi_B/\sqrt{2}$, $\tilde{x} = \sqrt{2} x/\xi_B$ and $\xi_B^2 = 1/2n_B \tilde{g}_B m_B$ the coherence length in the 1D Bose gas. Further, $G$ is the strength of the induced interaction given by

$$G = \left(\frac{g_{BF}}{2\pi l_\perp^2}\right)^2 \sqrt{2} n_B m_B \xi_B. \quad (5)$$

The remaining integral yields $\pi e^{-|\tilde{x}|}$. Hence, the real space induced interaction between fermions located at sites $j$ and $l$, with spacing $x = x_j - x_l$, is

$$V_{\text{ind}}(j, l) = -G \ e^{-\sqrt{2}|x_j - x_l|/\xi_B}. \quad (6)$$

It follows from (6) that one can tune both the strength and the range of the induced interaction by changing the boson–fermion and boson–boson interaction strength, which is a very attractive feature of this setup. For a spin-polarised 2D Fermi gas interacting via density oscillations in a 3D BEC, this flexibility enables one to tune the critical temperature of a topological superfluid to be close to the maximum value allowed by Kosterlitz–Thouless theory [32]. Note that whereas the induced interaction mediated by a 3D BEC gas is of the Yukawa form $\propto e^{-\sqrt{2}|x_j - x_l|/\xi_B}/|x|$ [32], the interaction (6) mediated by a 1D Bose gas falls of as $\sim e^{-\sqrt{2}|x_j - x_l|/\xi_B}$. We shall see that, in addition to the next-nearest neighbour hopping, the missing $1/|x|$ factor in the induced interaction is another crucial ingredient for producing states with a Chern number larger than one. In the following, we take the temperature to be zero.

## 4. Superfluid pairing

The induced interaction between the fermions is attractive, which means that they can form Cooper pairs. We shall investigate this using mean-field BCS theory. The validity of this approach is not obvious, since there is no true long range order corresponding to a broken continuous symmetry for our 1D system due to the Mermin–Wagner–Hohenberg theorem [34, 35]. However, mean-field theory in general works better when the interaction is long range since each particle interacts with a cloud of particles in distance. Because the induced interaction (6) is long range for large coherence lengths, $\xi_B/d \gg 1$, we expect that BCS theory will be qualitatively correct for the present system. This expectation is in part supported by recent work concerning an exactly solvable 1D model with a $1/r$ interaction, which was shown to support a bulk energy gap indicating order on a macroscopic scale [36]. Moreover, since the interaction in the present case can be made effectively constant over several lattice sites, we expect our system to be ordered on a macroscopic scale compared to the system size, thereby making BCS theory qualitatively correct.

The BCS Hamiltonian describing pairing of identical fermions with opposite momentum is

$$H_{\text{BCS}} = \sum_k \begin{bmatrix} c_k^\dagger & c_{-k} \end{bmatrix} \begin{bmatrix} \xi_k & \Delta_k \\ \Delta_k & -\xi_k \end{bmatrix} \begin{bmatrix} c_k \\ c_{-k}^\dagger \end{bmatrix}, \quad (7)$$

with $\xi_k = \varepsilon_k - \mu$ and $\mu$ being the chemical potential. The gap is given by

$$\Delta_k = -N^{-1} \sum_q W_{\text{ind}}(k, q) \langle c_q c_{-q} \rangle$$

$$= -\frac{1}{N} \sum_q W_{\text{ind}}(k, q) \frac{\Delta_q}{2E_q}, \quad (8)$$

where $N$ is the number of lattice sites, and $E_k = (\xi_k^2 + \Delta_k^2)^{1/2}$ is the usual energy dispersion of the quasiparticle eigenstates. Since the fermions are identical, the Pauli principle dictates that the pairing has the 'p-wave' symmetry $\Delta_k = -\Delta_{-k}$. The form of the effective interaction entering the gap equation is given by $W_{\text{ind}}(k, q) = 1/2 \sum_j [\cos((k-q)x_j) - \cos((k+q)x_j)] \tilde{V}_{\text{ind}}(0, j)$ where the cosines arise from the Fermi antisymmetry $\langle c_q c_{-q} \rangle = -\langle c_{-q} c_q \rangle$. We assume without loss of generality that the gap is real. The filling fraction $0 \leqslant n \leqslant 1$ of the lattice is

$$n = \frac{N_F}{N} = \frac{1}{2}\left(1 - \frac{1}{N} \sum_k \frac{\xi_k}{E_k}\right), \quad (9)$$

and the BCS ground state energy is $E_{\text{BCS}} - \mu N_F = \sum_k [\xi_k - E_k + |\Delta_k|^2/2E_k]/2$, where $N_F$ the number of fermions in the lattice.

For a vanishing next-nearest neighbour hopping, $t_2 = 0$, the system has a filling fraction symmetry between $n$ and $1 - n$ around $n = 1/2$. Specifically, any solution with $n$, $\mu$ and $\Delta_k$ has a corresponding solution $1 - n$, $-\mu$ and $\Delta_{k+\pi/d}$. For nonzero $t_2$ this symmetry is broken. We will see this explicitly below in the phase diagrams.

## 5. Symmetries and topology

As mentioned in section 1, the Kitaev chain belongs to the BDI Cartan class due to the presence of an intrinsic particle–hole symmetry and a time-reversal symmetry, both squaring to $+1$. As a result the topological index is given by the Chern number

$$C = \frac{i}{\pi} \int_{-\pi/d}^{\pi/d} dk \langle e_k^- | \partial_k | e_k^- \rangle$$

$$= \frac{1}{2\pi} \int_{-\pi/d}^{\pi/d} dk \frac{\xi_k \partial_k \Delta_k - \Delta_k \partial_k \xi_k}{\xi_k^2 + \Delta_k^2} = \int_{-\pi/d}^{\pi/d} dk \frac{\partial_k \theta_k}{2\pi}, \quad (10)$$

where $|e_k^-\rangle$ is the lowest eigenstate of $V^\dagger \mathcal{H}(k) V$, with $\mathcal{H}(k) = \begin{bmatrix} \xi_k & \Delta_k \\ \Delta_k & -\xi_k \end{bmatrix}$ and $V = \frac{1}{\sqrt{2}} \begin{bmatrix} 1 & 1 \\ i & -i \end{bmatrix}$. We apply the unitary transformation matrix $V$ to $\mathcal{H}(k)$ in order to make the eigenvector $|e_k^-\rangle$ smooth for all $k$ [37]. In (10), $\theta_k = \arctan(\Delta_k/\xi_k)$ is the polar angle of the vector $\mathbf{h}(k) = (\Delta_k, 0, \xi_k)$ with the $z$-axis, where $\mathcal{H}(k) = \mathbf{h}(k) \cdot (\sigma_x, \sigma_y, \sigma_z)$. Equation (10) therefore explicitly demonstrates the well-known result that the topological invariant given by the Chern number is simply the same as the winding number of $\mathbf{h}$ for a 1D system.

We can obtain the Chern number by simply examining the behaviour of $\Delta_k$ and $\xi_k$ as a function of $k$, without having to actually evaluate the integral in (10). The vector $\mathbf{h}(k) = (\Delta_k, 0, \xi_k)$ is confined to the $xz$-plane and it starts and ends parallel to the $z$-axis since $\Delta_{-\pi/d} = \Delta_{\pi/d} = 0$. Using the symmetries, $\Delta_k = -\Delta_{-k}$ and $\xi_k = \xi_{-k}$, we can then obtain the winding number simply by counting how many times $\mathbf{h}(k)$ crosses the $x$-axis for $k > 0$. Now, $\mathbf{h}(k)$ crosses the $x$-axis when $\xi_k = 0$. Assuming $\Delta_k > 0$, the crossing is in the clockwise direction with increasing $k$ when $\partial_k \xi_k < 0$ and is in the counterclockwise direction when $\partial_k \xi_k > 0$. On the other hand, if $\Delta_k < 0$, the crossing is in the clockwise/counterclockwise direction with increasing $k$ for $\partial_k \xi_k > 0$ and $\partial_k \xi_k < 0$, respectively. A clockwise crossing increases the winding number by one, whereas a counterclockwise

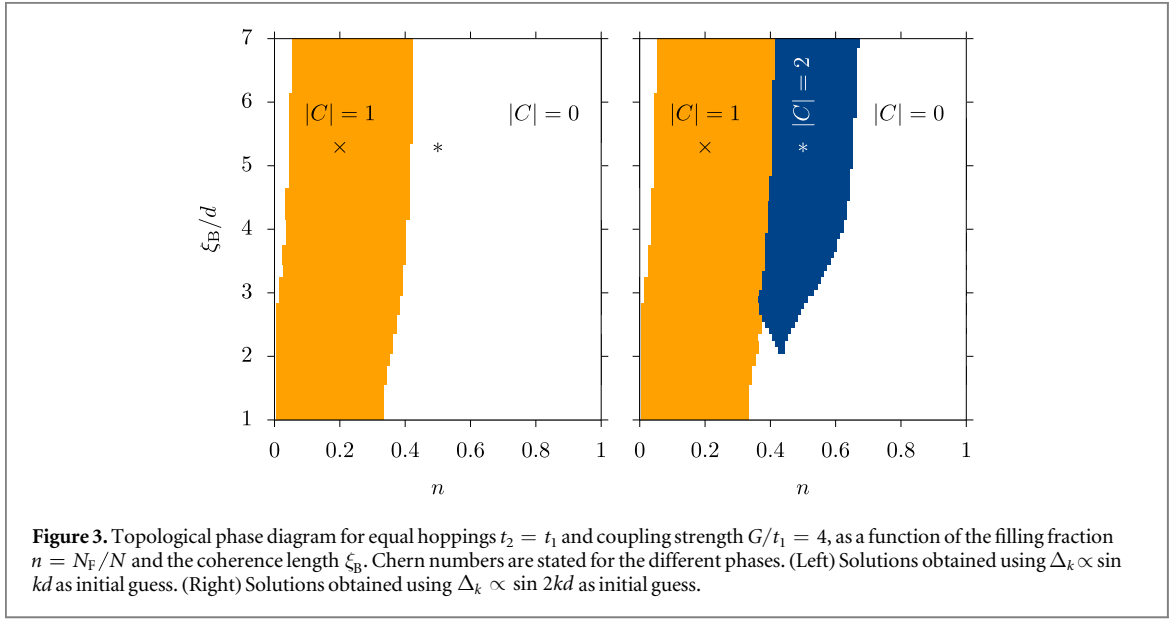

**Figure 3.** Topological phase diagram for equal hoppings $t_2 = t_1$ and coupling strength $G/t_1 = 4$, as a function of the filling fraction $n = N_F/N$ and the coherence length $\xi_B$. Chern numbers are stated for the different phases. (Left) Solutions obtained using $\Delta_k \propto \sin kd$ as initial guess. (Right) Solutions obtained using $\Delta_k \propto \sin 2kd$ as initial guess.

crossing decreases it by one. This finally gives the formula

$$C = -\sum_{\substack{k_n > 0 \\ \xi_{k_n}=0}} \text{sgn}(\Delta_{k_n} \partial_k \xi_{k=k_n}) \quad (11)$$

for the Chern number. Here $k_n$ are the solutions to $\xi_k = 0$ and $\text{sgn}(x)$ is the sign function.

We can now see why it is important to have next-nearest neighbour hopping in order to get a Chern number larger than 1. For nearest neighbour hopping only, $\xi_k = 0$ has at most one solution for $k > 0$ and the winding number is therefore $C = 0, \pm 1$ from (11). With next-nearest neighbour hopping $t_2 \neq 0$ included, we can have up to two zero points of $\xi_k$ for $k > 0$. Since $\partial_k \xi_k$ has opposite signs at the two zero points, we can obtain a winding number $C = \pm 2$ only if the pairing $\Delta_k$ changes sign between the two points. Such a pairing indeed occurs and is functionally similar to $\Delta_k \sim \sin 2kd$ as we will see in the numerical analysis below. To summarise, we need nonzero next-nearest neighbour hopping $t_2$ and a high filling fraction such that there are two pairs of zeroes of $\xi_k$ and a pairing $\Delta_k \sim \sin 2kd$. For a given interaction strength $G$, we will therefore search for values of $(\xi_B, n)$ where the pairing $\Delta_k \sim \sin 2kd$ is favourable.

## 6. Numerical results

We now discuss our results obtained by solving (8) and (9) numerically. We look for solutions with Chern number $|C| = 1$ or $|C| = 2$ by using the pairing functions $\Delta_k \propto \sin kd$ and $\Delta_k \propto \sin 2kd$ as initial guesses, respectively.

In figure 3, we show the resulting zero temperature phase diagrams obtained by varying the BEC coherence length $\xi_B$ and the filling fraction $n$, for a fixed coupling strength $G/t_1 = 4$ and a fixed next-nearest neighbour hopping $t_2/t_1 = 1$. Physically, this corresponds to varying the boson–boson interaction strength $g_B$ and the filling fraction $n$, while keeping the Bose–Fermi coupling strength $g_{BF}$ and the BEC density $n_B$ fixed. Consider first the left phase diagram, which is obtained by using $\Delta_k \propto \sin kd$ as an initial guess. First, note that the phase diagram is highly asymmetrical around $n = 1/2$, consistent with the observation that a nonzero next-nearest neighbour hopping breaks the $n \leftrightarrow 1 - n$ symmetry. Second, we see that there is a topologically non-trivial region with Chern number $|C| = 1$ for intermediate filling fractions. To understand this in more detail and to prepare ourselves for the ensuing analysis of the phase with Chern number 2, we plot in figure 4 $\Delta_k$ and $\xi_k$ for the two specific points indicated by the symbols $\times$ and $*$ in the phase diagram figure 3 (left). For the point $\times$, corresponding to $(n, \xi_B/d) = (0.2, 5)$, figure 4 (left) shows that there is only one solution to $\xi_k = 0$ for $k > 0$, and we get from (11) a Chern number $C = -1$ since $\Delta_k > 0$ and $\partial_k \xi_k > 0$ at that point. The point $*$ corresponds to a larger filling fraction $(n, \xi_B/d) = (0.5, 5)$, and in this case $\xi_k = 0$ has two solutions as can be seen from figure 4 (right). Since $\Delta_k > 0$ for both solutions whereas $\partial_k \xi_k$ has opposite signs, the Chern number is zero.

Consider next the phase diagram shown figure 3 (right), which is obtained using the initial guess $\Delta_k \propto \sin 2kd$. We see that there is now a large region with a Chern number $|C| = 2$. To analyse this, we plot in figure 4 (right) $\Delta_k$ and $\xi_k$ for the point marked by $*$ in the phase diagram figure 3 (right) corresponding to $(n, \xi_B/d) = (0.5, 5)$. Again, $\xi_k = 0$ has two solutions with opposite sign of the slopes $\partial_k \xi_k$. Contrary to the self-consistent solution above obtained from using $\Delta_k \propto \sin kd$ as an initial guess, $\Delta_k$ now has opposite signs at the two solutions, and it follows from (11) that the Chern number is $C = -2$. Using the bulk-boundary principle, we conclude that this phase supports two mutually orthogonal Majorana modes at its ends.

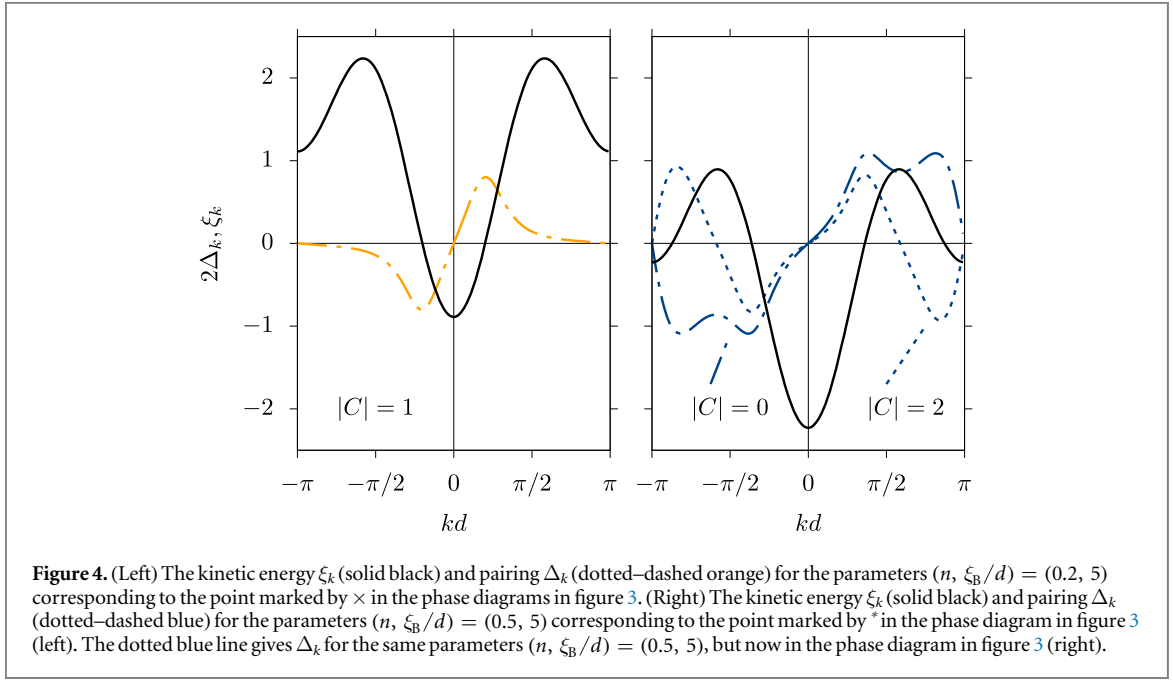

**Figure 4.** (Left) The kinetic energy $\xi_k$ (solid black) and pairing $\Delta_k$ (dotted–dashed orange) for the parameters $(n, \xi_B/d) = (0.2, 5)$ corresponding to the point marked by × in the phase diagrams in figure 3. (Right) The kinetic energy $\xi_k$ (solid black) and pairing $\Delta_k$ (dotted–dashed blue) for the parameters $(n, \xi_B/d) = (0.5, 5)$ corresponding to the point marked by * in the phase diagram in figure 3 (left). The dotted blue line gives $\Delta_k$ for the same parameters $(n, \xi_B/d) = (0.5, 5)$, but now in the phase diagram in figure 3 (right).

Figure 3 (right) shows that the phase with Chern number $|C| = 2$ is stable only for coherence lengths $\xi_B/d \gtrsim 2$. This can be understood from the fact that $\Delta_k \sim \sin kd$ corresponds to predominantly nearest neighbour pairing, whereas $\Delta_k \sim \sin 2kd$ corresponds to predominantly next-nearest neighbour pairing. Thus, in order to obtain a Chern number number larger than 1, we need a significant next-nearest neighbour pairing, which in turn requires that the range of the induced interaction $\xi_B$ be on the order of twice the lattice spacing or larger.

Comparing figure 3 (left) and (right), we see that there is a large region of the phase diagram in which two possible self-consistent solutions exist corresponding to phases with Chern numbers $C = 0$ and $|C| = 2$. Interestingly, the energy of the two phases turn out to differ only by a few percent, with the $C = 0$ being the lower one. Such a small energy difference means that the two phases are practically degenerate, and there is therefore a large probability that the system ends up in the $|C| = 2$ phase when cooled down.

To illustrate the significance of the next-nearest neighbour hopping, we plot in figure 5 the phase diagram obtained for $G/t_1 = 4$ and $t_2/t_1 = 0.63$. The smaller next-nearest neighbour hopping has two effects on the system. The dominant one is that $|C| = 2$ phase only exists for even larger coherence lengths, $\xi_B/d \gtrsim 5$. The reason is that for a smaller $t_2$, the zeros of $\xi_k$ are closer in $k$-space. In order to a get a Chern number $|C| = 2$, the gap then has to change sign faster, which requires higher harmonics of $\Delta_k$. Such higher harmonics are generated by pairing between fermions with large spacial separation beyond next-nearest neighbour, which in turn requires a longer range interaction and thus a larger $\xi_B/d$. Secondly, the $|C| = 2$ phase appears at higher filling fractions, $n \approx 0.6$. This is because the filling fraction now needs to be higher to create two zeroes of $\xi_k$.

## 7. Discussion

All the ingredients necessary for realising the proposed system have been realised experimentally with cold atom systems. Periodically modulated optical lattices as well as Bose–Fermi mixtures have been created by several groups, and this includes, in particular, multicomponent systems in 1D [38] with a species selective potential [39, 40]. This makes the experimental realisation of the system promising, which is a strong feature of our proposal. There are however two caveats about our proposal: one should be able to cool the system sufficiently, and fluctuations away from mean-field can in principle modify some of the results we found above. Nevertheless, we speculate that the topological features should be fairly robust since they are determined by invariants which take on integer values. In addition, fluctuation effects are suppressed for a long range interaction.

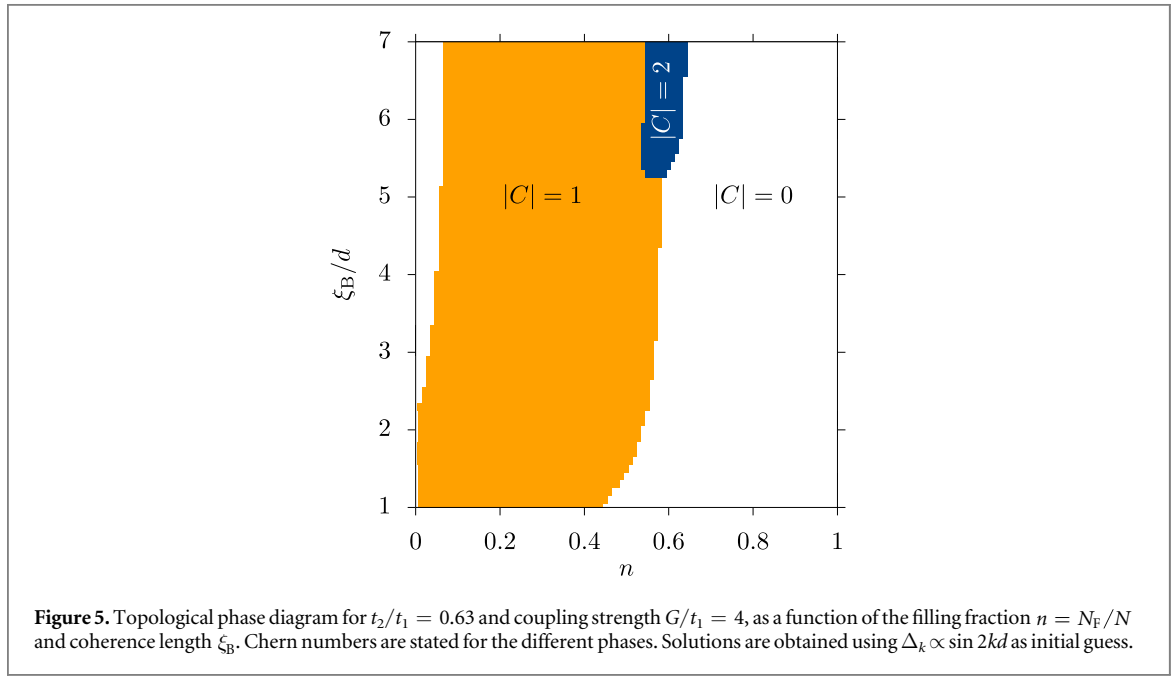

**Figure 5.** Topological phase diagram for $t_2/t_1 = 0.63$ and coupling strength $G/t_1 = 4$, as a function of the filling fraction $n = N_F/N$ and coherence length $\xi_B$. Chern numbers are stated for the different phases. Solutions are obtained using $\Delta_k \propto \sin 2kd$ as initial guess.

We note that a harmonic trapping potential, which is present in the vast majority of cold atom experiments, can stabilise the presence of a BEC in 1D. Whereas the low-energy density-of-states (DOS) diverges for a 1D homogeneous system, it is a constant for a 1D harmonically trapped system. It follows that in terms of the DOS, a 1D trapped system behaves as a 2D homogeneous system, for which there is a quasi-condensate at finite temperature with algebraic order, which turns into a proper BEC with long range order at zero temperature. This line of reasoning gives further support to our use of Bogoliubov theory for the bosons.

The periodic driving of the lattice will in general heat the system, but this will become significant only after a certain time scale $t_h$—the so-called heating time. The effective Hamiltonian derived from a high frequency expansion, in which the ratio between the next-nearest neighbour hopping and the nearest neighbour hopping is enhanced, is valid to describe the physics within the time period $t \ll t_h$. In a typical cold atomic experiment involving an optical lattice, the heating time increases exponentially in terms of the driving frequency (for driving frequencies smaller than the band gap) and can be as long as the lifetime of the experiments [41]. This effect has been used to minimise the heating, and insignificant heating was indeed observed in experiments using Floquet engineering [26, 29, 30].

We end by briefly discussing the case where the 1D lattice for the fermions is immersed in a 3D BEC. Our numerical calculations show that it is much more difficult to find phases with Chern numbers larger than 1 compared to the pure 1D system discussed above. Their energy is furthermore much higher than phases with Chern number 0 or 1, again in contrast to the pure 1D case. The reason is that the induced interaction between the fermions is of the Yukawa form $\propto \exp(-\sqrt{2}\,|x|/\xi_B)/|x|$ when the BEC is 3D; the extra factor $1/|x|$ compared to the pure 1D interaction (6) suppresses the next-nearest neighbour pairing compared to nearest neighbour pairing irrespective of the value of $\xi_B$.

## 8. Conclusions

In conclusion, we considered a spin-polarised Fermi gas in a 1D lattice immersed in a 1D homogeneous Bose gas. The fermions interact attractively via the exchange of density fluctuations in the Bose gas, which gives rise to pairing. We derived a simple formula for the Chern number of the superfluid phase, which was used to demonstrate that the fermions can realise a topological superfluid with Chern number $\pm 2$, provided that the next-nearest neighbour hopping for the fermions is significant and the induced interaction is sufficiently long range. This phase was shown to be stable in a large region of the phase diagram as a function of the filling fraction of the fermions and the coherence length of the bosons. Atomic gases provide a promising system to realise our proposed setup using well established experimental techniques.

## Acknowledgements

We acknowledge several useful discussions with Jonatan Melkær Midtgaard. GMB wishes to acknowledge the support of the Danish Council of Independent Research–Natural Sciences via Grant No. DFF-4002-00336.